\documentclass[UKenglish,texlive=2011,txfonts]{article}
\pdfoutput=1

\usepackage{defs}
\usepackage{graphicx}
\graphicspath{{../figures/}}

\title{`That looks weird' --- evaluating citizen scientists'
ability to detect unusual features in ATLAS images of collisions
at the Large Hadron Collider}

\begin{document}
\author[a]{A.J~Barr}
\author[b]{A.~Haas}
\author[a,c]{C.W.~Kalderon}
\affil[a]{Department of Physics, University of Oxford, Oxford, UK}
\affil[b]{Department of Physics, NYU, New York, USA}
\affil[c]{Department of Physics, University of Lund, Sweden}

\maketitle

\abstract{%
Using data from the \url{HiggsHunters.org} project 
we investigate the ability of non-expert citizen scientists to 
identify long-lived particles, and other unusual features,
in images of LHC collisions recorded by the ATLAS experiment.
More than \nvolunteers{} volunteers from \ncountries{} countries
participated, classifying \nclicks{} features of interest
on about \nimagesviewed{} distinct images.
We find that the non-expert volunteers are capable
of identifying the decays of long-lived particles
with an efficiency and fake-rate comparable to that
of the ATLAS algorithms.
Volunteers also picked out events with unexpected features, including
what appeared to be an event containing a jet of muons.
A survey of volunteers indicates a high level of scientific engagement
and an appetite for further LHC-related citizen science projects.
}

\section{Introduction}
\label{sec:intro}

The Large Hadron Collider is arguably the highest profile scientific project of our time.
The discovery of the Higgs boson~\cite{HIGG-2012-27,CMS-HIG-12-028} has been the scientific highlight to date.
The accelerator continues to be the subject of much media attention as searches for other new particles continue.

\begin{figure}
\centering
\includegraphics[width=0.7\linewidth]{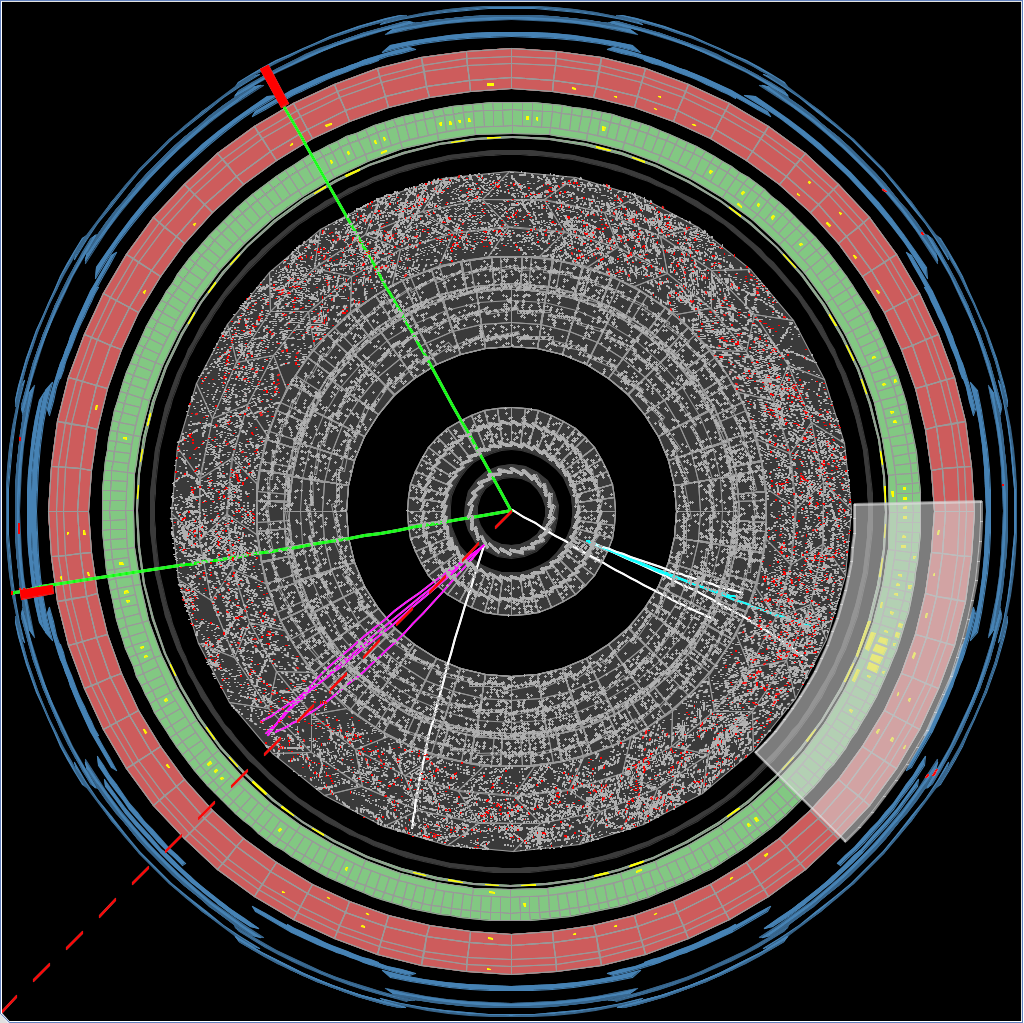}
\caption{\label{fig:exampleimage}
An example ATLAS detector image presented to citizen scientists. 
This image contains two off-centre vertices, 
each visible as a vee-like structure, 
at about 4 o'clock and 7 o'clock, a little distance from the center of the image.
The image was generated from a computer simulation of the process $H \to \phi + \phi$. 
The green lines emanating from the centre 
indicate the reconstructed muon and antimuon used to select the event. 
The red dotted line indicates the direction of the missing momentum 
transverse to the beam.
}
\end{figure}

Matching this cutting edge science with the public's curiosity to understand it can present a challenge. 
The particles themselves are invisible. Most decay a tiny fraction of a second after their creation, and can only be detected and reconstructed 
using large dedicated detectors assembled over decades by large international collaborations.

Despite these difficulties, there is a strong drive within science policy to 
get the public more involved in not just reading about science,
but actually performing it.
Citizen science projects --
which directly involve the public in the scientific process --
represent an ideal vehicle for meaningful engagement with a large community.

Non-expert citizen scientists have previously been shown 
to be good classifiers of images~\cite{zooniverse_papers}. 
They are also capable of spotting unusual objects in images
including unexpected galaxy features~\cite{Lintott11102009}.
Through the Galaxy Zoo~\cite{galaxyzoo} project alone, citizen scientists 
have contributed to the results of 48 scientific 
papers~\cite{zooniverse_papers}.
The present study evaluates
the extent to which analysis by non-expert
citizen scientists might also be possible at the LHC.

Previously the public has been invited to contribute to 
CERN's science by donating idle time on their computer to help simulate 
proton-proton collisions~\cite{lhcathome,atlasathome}.
This helps the scientific endeavour but with the caveat that
the individual member of the public is effectively more of a provider 
of computing resource than an active researcher.
More direct involvement in the research has previously been
restricted to the relatively small fraction of the public that has a high level of 
computing coding skills.
Such individuals have been able to directly analyse data from the 
ATLAS, CMS, LHCb and ALICE experiments 
via the CERN opendata portal~\cite{opendata}.
The Kaggle project~\cite{kaggle} in which members of the public were challenged 
to use machine learning to identify Higgs boson events
also demanded a high level of coding expertise.

The HiggsHunters project is, to the best of our knowledge, the first to allow the 
non-expert general public a direct role in searching for new particles at the 
LHC. Though new in this context, 
it is revival of a long-established technique --
before the invention of electronic particle detectors 
eye-scanning images of particle tracks by trained technicians  
was the standard analysis method.

\section{Physics model and classification task}
\label{sec:model}

\begin{mybox}[tb]
\centering\begin{minipage}{0.9\linewidth}
\begin{framed}
\begin{center}\textbf{Target bosons}\end{center}
\vskip 2mm
The theories of most interest to us 
predict the existence of new particles $\phi$ which are not in the 
Standard Model and which 
have not yet been observed experimentally.
In such theories the usual Higgs boson $H$, after it is created, 
would most often decay as predicted by the Standard Model,
however a fraction of the time it would decay into 
the new particles:
\[
H \to \phi + \phi.
\]
The new particles $\phi$ interact with the Standard Model only very weakly.
This weak coupling means they have a slow decay rate,
and hence a relatively long lifetime on the particle scale
-- typically of order nanoseconds.
They can therefore travel a macroscopic distance,
perhaps tens of centimetres, before themselves decaying.
\end{framed}
\end{minipage}
\end{mybox}

For the \url{HiggsHunters.org} project, 
a task was required which lent itself well to the strengths
of non-expert citizen scientists -- in particular their abilities
to classify images, and to spot unusual features.
The task selected was that of identifying new particles $\phi$ 
--- our `\target{}' bosons 
--- as they decay within the ATLAS detector~\cite{PERF-2007-01}.
Such particles are predicted in theories in which an additional scalar
mixes weakly with the Standard Model Higgs boson~\cite{Strassler:2006ri}.

The selected processes therefore generate a signature that is fairly easily 
identifiable by eye (\cref{fig:exampleimage}), and which 
citizen scientists might therefore 
be competitive with a standard reconstruction algorithm.
The fact that no $\phi$ boson had been observed to date was also a very desirable 
feature. Unambiguous observation of evidence for these new particles
would be a very significant scientific discovery, comparable to 
that of the discovery of the Higgs boson itself.
The high impact of a potential discovery 
meets the important motivating feature of citizen science projects that 
the volunteers have a real opportunity of discovering something previously 
unknown to science.

\begin{mybox}[tb]
\centering\begin{minipage}{0.9\linewidth}
\begin{framed}
\begin{center}\textbf{Decay modes}\end{center}
\vskip 2mm
The most likely decay mode of the new $\phi$ boson depends on its mass $m_\phi$.
If $m_\phi$ is at least twice the mass of the $b$ quark,
the \target{} boson will mostly decay 
to a bottom quark $b$ and its anti-particle $\bar{b}$
\[
\phi \to b + \bar{b}.
\]
If the mass is smaller, lying in the range $2m_\tau < m_\phi < 2m_b$,
then the dominant decay is to a $\tau$ lepton and its anti-particle
\[
\phi \to \tau^+ + \tau^-,
\]
where $m_\tau$ and $m_b$ are the mass of the $\tau$ lepton and the $b$
quark respectively. The $\tau$ leptons and $b$ quarks themselves
have rather short lifetimes -- and decay in of order picoseconds
typically to many charged particles, each of which will leave a 
distinctive track in the ATLAS detector.
\end{framed}
\end{minipage}
\end{mybox}

\section{Image selection}\label{sec:images}

The ATLAS experiment~\cite{PERF-2007-01} is positioned
at one of the four interaction points in the LHC.
It comprises a central tracking detector,
surrounded by electromagnetic and hadronic calorimeters, 
which are themselves surrounded by a dedicated muon detector.
Within the inner tracking detector, the paths of charge particles are bent
by the field from a superconducting solenoidal magnet.
A separate system of magnets provides a toroidal field
within the muon detector.

The Large Hadron Collider provides twenty million 
proton-proton bunch crossings per second, 
far more than is practical for ATLAS to record.
To reduce the data volume,
the ATLAS trigger algorithms~\cite{PERF-2011-02} select 
up to several hundred of those events each second
having identified features of interest such as muons, electrons,
or high energy jets of hadrons.

Further selection was required before presenting images to volunteers.
Given the anticipated number of volunteers (of the order of $10^4$ to $10^5$)
the likely number of classifications per volunteer (anticipated to be on average ten
but with a long tail of enthusiastic classifiers)
and the required level of redundant classification to allow for robust analysis
the desired number of images was in the range $10^4$ to $10^5$ events.

It was possible to select the required number of events, and
simultaneously enrich them in events likely to contain Higgs bosons, 
by pre-selecting those events containing a muon and an antimuon
with invariant mass consistent the mass of the $Z$ boson.
Such events are consistent with the $Z$ boson decay process $Z \to \mu^+ + \mu^-$.
Events containing a $Z$ boson have an increased probability
of also containing a Higgs boson, because
virtual $Z$ bosons may emit Higgs boson through the process 
$Z^* \to Z + H$
known as `Higgs-strahlung'. 

ATLAS uses a right-handed coordinate system with its origin at the nominal interaction point (IP) in the centre of the detector and the $z$-axis along the beam pipe. The $x$-axis points from the IP to the centre of the LHC ring, and the $y$-axis points upward. Cylindrical coordinates $(r,\phi)$ are used in the transverse plane, $\phi$ being the azimuthal angle around the $z$-axis. The pseudorapidity is defined in terms of the polar angle $\theta$ as $\eta=-\ln\tan(\theta/2)$.

To enrich the data sample in Higgs-strahlung events over $Z \to \mu^+ + \mu^-$ alone, 
the reconstructed $Z$ boson was required to have a transverse momentum, \pt, greater than 60~\GeV,
a criterion which retains about 60\% of the $Z+H$ events but just 5\% of the \zjets background. 
A further 50\% of the \zjets background is removed through requiring 
that the missing transverse momentum be larger than $40\,$GeV. 
This requirement enhances the signal-to-background ratio, 
since the $b$-quark and $\tau$-lepton decays in the signal events 
often lead to a transverse momentum imbalance.

Data were selected from the 2012 data-taking period, during the period April to December, 
during which time the LHC was colliding protons against protons 
at a centre-of-mass energy of $8 \TeV$. 
The data set selected corresponds to an integrated luminosity of about 12\,fb$^{-1}$,
and results in approximately 60,000 candidate $Z \to \mu^+ + \mu^-$ events 
of which around 60 are expected to feature a Higgs boson. 

The ability of volunteers to identify the off-centre vertices
was calibrated using test images which showed 
Monte Carlo simulations
of the process $H \to \phi+\phi$ of interest. 

Several different $\phi$ boson
masses $m_\phi$ and average lifetimes $\tauphi$ were investigated. 
A summary of the different processes considered can be found in \cref{tab:images}.

\begin{table*}
\centering
\renewcommand\arraystretch{1.35}
\setlength{\tabcolsep}{1mm}
\begin{tabular}{l c c c c r r}
\toprule
\multirow{2}{*}{Process} & \multicolumn{4}{c}{$\phi$ particle properties (where relevant)} & \multirow{2}{*}{Events} & \multirow{2}{*}{$\geq 3$ views} \\
                         & mass $m_\phi$ & $\phi$ decays to & $c\tauphi$ & flight distance $\gamma \beta c \tauphi$ & \\ 
\midrule
$Z+H$ & 8 GeV  & $\tau^+ + \tau^-$ & 1 mm   & 7.7 mm   & 3103 & 605 \\
$Z+H$ & 8 GeV  & $\tau^+ + \tau^-$ & 10 mm  & 77 mm  & 3119 & 988 \\
$Z+H$ & 8 GeV  & $\tau^+ + \tau^-$ & 100 mm & 770 mm & 1277 & 197 \\
$Z+H$ & 20 GeV & $b + \bar{b}$     & 1 mm   & 3.0 mm   & 3012 & 1240 \\
$Z+H$ & 20 GeV & $b + \bar{b}$     & 10 mm  & 30 mm  & 3094 & 1962 \\
$Z+H$ & 20 GeV & $b + \bar{b}$     & 100 mm & 300 mm & 1341 & 485 \\
$Z+H$ & 50 GeV & $b + \bar{b}$     & 1 mm   & 0.75 mm & 2954 & 1183 \\
$Z+H$ & 50 GeV & $b + \bar{b}$     & 10 mm  & 7.5 mm   & 3121 & 1894 \\
$Z+H$ & 50 GeV & $b + \bar{b}$     & 100 mm & 75 mm  & 1294 & 663 \\
$Z \rightarrow \mu \mu$ & \multicolumn{4}{c}{---}   &  374 & 321 \\
\midrule
Data & \multicolumn{4}{c}{---}                     & 62278 & 13955 \\
Data (debug) & \multicolumn{4}{c}{---}             & 207   & 178 \\
\midrule
\textbf{Total} & \multicolumn{4}{c}{---}   & \textbf{85174} & \textbf{23522} \\

\bottomrule
\end{tabular}
\caption{Properties of the simulated images.
The flight distance $\gamma \beta c \tauphi$ is that for $\phi$ particles produced 
from an at-rest Higgs boson, i.e. with a total energy of $m_H / 2$. 
The column headed ``$\geq 3$ views'' is the number of events that have had at least 
three citizen scientists interact with them for each of the three image projections. 
Where a citizen scientist examines an image but does not click, this is not counted in the above.
ATLAS data were sourced from two different streams -- the usual physics stream,
and the `debug' stream which contains events that caused problems during reconstruction.
\label{tab:images}}
\end{table*}

Since we are looking for decays away from the proton-proton interactions,
an interesting question is the average distance which the $\phi$ bosons 
can be expected to travel.
The distance can be calculated using a standard relativistic calculation.
The $\phi$ boson's average lifetime in its own rest frame is $\tauphi$.
Different lifetimes were selected by choosing values of $c\tauphi$
where $c$ is the speed of light.
When moving at speed $\beta$ relative to $c$
the lifetime of the $\phi$ boson will be increased (time dilated) 
by the relativistic Lorentz factor
\[
\gamma = \frac{1}{\sqrt{1-\beta^2}},
\]
meaning that the average distance travelled will be $\beta c \gamma \tauphi$.

The speed of the $\phi$ boson in the rest frame of the Higgs boson can 
also be calculated, this time using a relativistic energy calculation. 
The energy available in the Higgs boson rest frame is $m_H c^2$ 
so the energy of each $\phi$ boson from its decay is $m_H c^2/2$.
Since the energy of the $\phi$ bosons is given by $\gamma m_\phi c^2$,
the Lorentz gamma factor in that frame is given by
\[
\gamma = \frac{m_H}{2 m_\phi}.
\]
Thus we can calculate $\gamma$, $\beta$ can be calculated 
from the definition of $\gamma$ and from them find
the expected distance $\beta\gamma c\tauphi$ travelled
by the $\phi$ bosons in the Higgs boson rest frame,
also shown in \cref{tab:images}.
The $\phi$ boson lifetimes have been seleted such that
the average distances are in the order of millimetres to metres.

All images, whether simulation or data, were processed 
using the ATLAS reconstruction software~\cite{SOFT-2010-01}, 
with some additions as in Ref.~\cite{Aad:2015rba}. 
The most important features of that reconstruction 
for this purpose are the tracks in the inner detector from the interactions
of charged particles with detector elements.

For each event processed, three images are produced. 
Two of these are in the transverse plane perpendicular to the beam pipe, 
with one giving a full-detector view (\XY) and the other a view only of the 
inner tracking subdetectors (\XYzoom). The former allows for identification of 
`weird' features in events, while the latter provides a close-up of the part most relevant for 
the identification of off-centre vertices. 
In addition to these views, there is an additional zoomed-in view showing a projection along the beam pipe 
(\RZzoom), in principle allowing information in all three dimensions.
All images have the central parts of the detector magnified using a fisheye projection 
(see Appendix) in order to better see tracks and off-centre vertices.


As of October 2016 classifications had been performed by 32,288 citizen scientists, 
of whom 9,610 had created Zooniverse accounts.
New users are invited to create a Zooniverse account after their first five classifications, 
and periodically thereafter.
For those classifications made without Zooniverse accounts 
it is assumed that classifications from different IP 
addresses are distinct scientists, since without an account there is no way to correct for the same
individual classifying from different devices.

\section{Evaluation of Off-Centre Vertex Identification}
\label{sec:analysis:ability}

The question to be addressed is how effective the citizen scientists are at locating vertices. 
Their performance can compared against computer algorithms that were developed and used by the ATLAS collaboration to identify 
off-centre vertices~\cite{Aad:2015rba}, 
using images in which decays of new $\phi$ bosons have been simulated in $H \rightarrow \phi + \phi$ decays. 
There are two key indicators of performance: the rate at which volunteers correctly identify the off-centre vertices 
and the rate at which they incorrectly identify off-centre vertices that are not from such decays 
-- i.e. they are either image artefacts or result from some other (known and understood) process. 

Following common Zooniverse practice, 
each event is classified by $\sim$60 people, translating to $\sim$20 per image (three projections per image) -- 
including `classifications' where no image features were marked.
In what follows, only images where at least three 
citizen scientists have marked an off-centre vertex are considered. 

To aggregate classifications across citizen scientists, 
the \texttt{DBSCAN} clustering algorithm~\cite{DBSCAN} was used to find clusters of clicks.
It requires a clustering size parameter, $\epsilon$, in addition to the 2D co-ordinates of click positions. 
Further selections were imposed on the cluster candidates, requiring them to be formed of 
at least $N_\textrm{clicks}$ individual clicks,
to have contributions from at least a fraction $f_\textrm{clicks}$ of people who marked a vertex on the image,
and to be formed only from clicks where the vertex has been identified by the citizen scientist 
as featuring at least $N_\textrm{tracks}$ tracks (which can be zero in the case that no number was provided by a citizen scientist). 
Once a parameter choice has been made and a set of clusters formed for each image, the refined clusters 
are compared to the known decay positions 
of the $\phi$ particles in the simulation. 
If a cluster is closer than 25 pixels (images are $1024\times1024$ pixels) to the nearest decay position, 
it is considered to be a correct identification and both cluster and true decay position are removed from consideration, 
before repeating the matching for remaining clusters. 
Any cluster not matched to a true decay position in this way is considered a `fake' cluster. 
Then the efficiency is $\sum_{N_\textrm{events}} N_\textrm{matched clusters} / 2 \times N_\textrm{events}$ 
and the fake rate is $\sum_{N_\textrm{events}} N_\textrm{unmatched clusters} / N_\textrm{events}$, since there are two true vertices per image. 

\begin{figure}
\centering
\begin{subfigure}[b]{0.49\textwidth}
        \includegraphics[width=\textwidth]{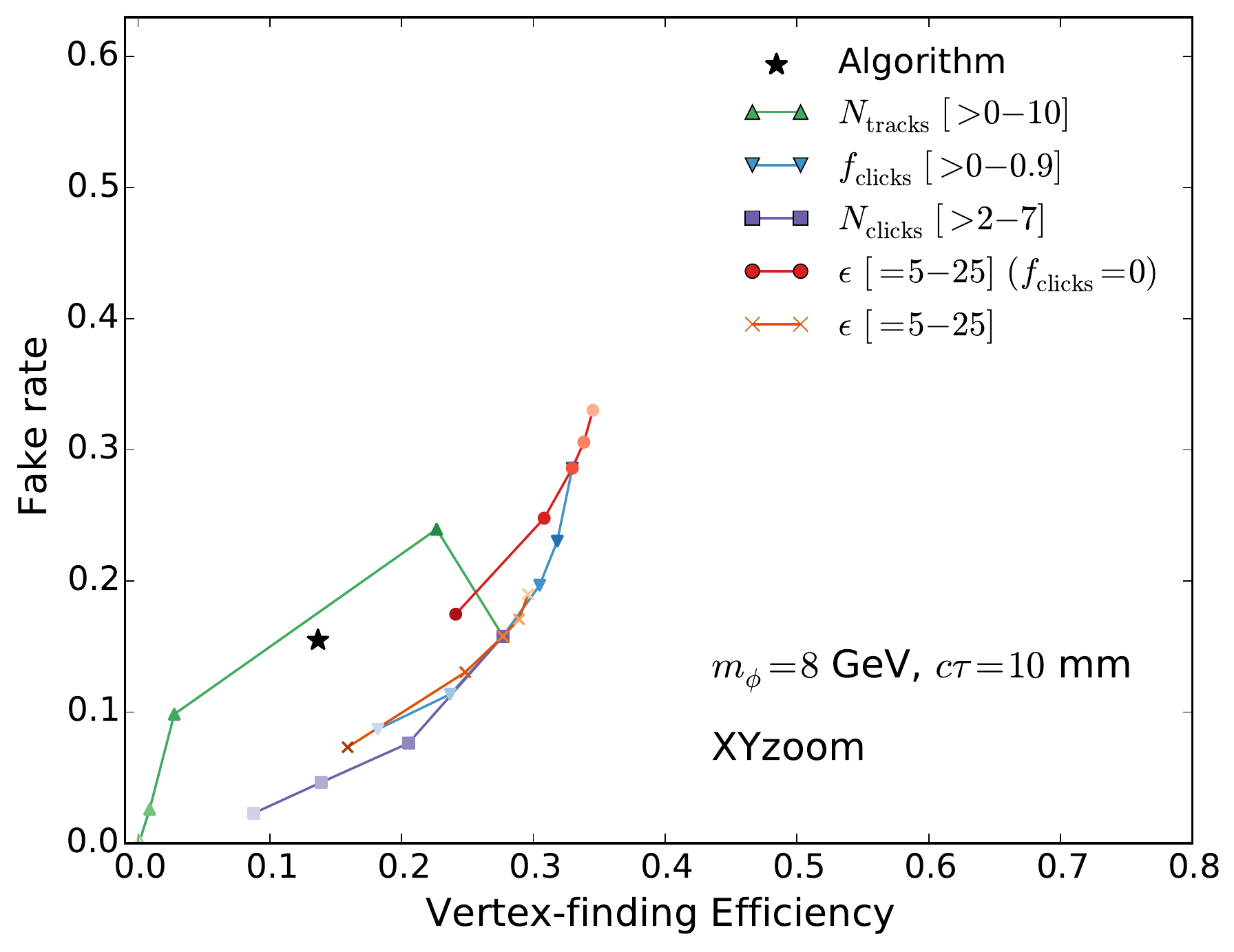}
        \caption{$m_\phi$ = 8 GeV, $c\tauphi$ = 10 mm}
\end{subfigure}
\begin{subfigure}[b]{0.49\textwidth}
        \includegraphics[width=\textwidth]{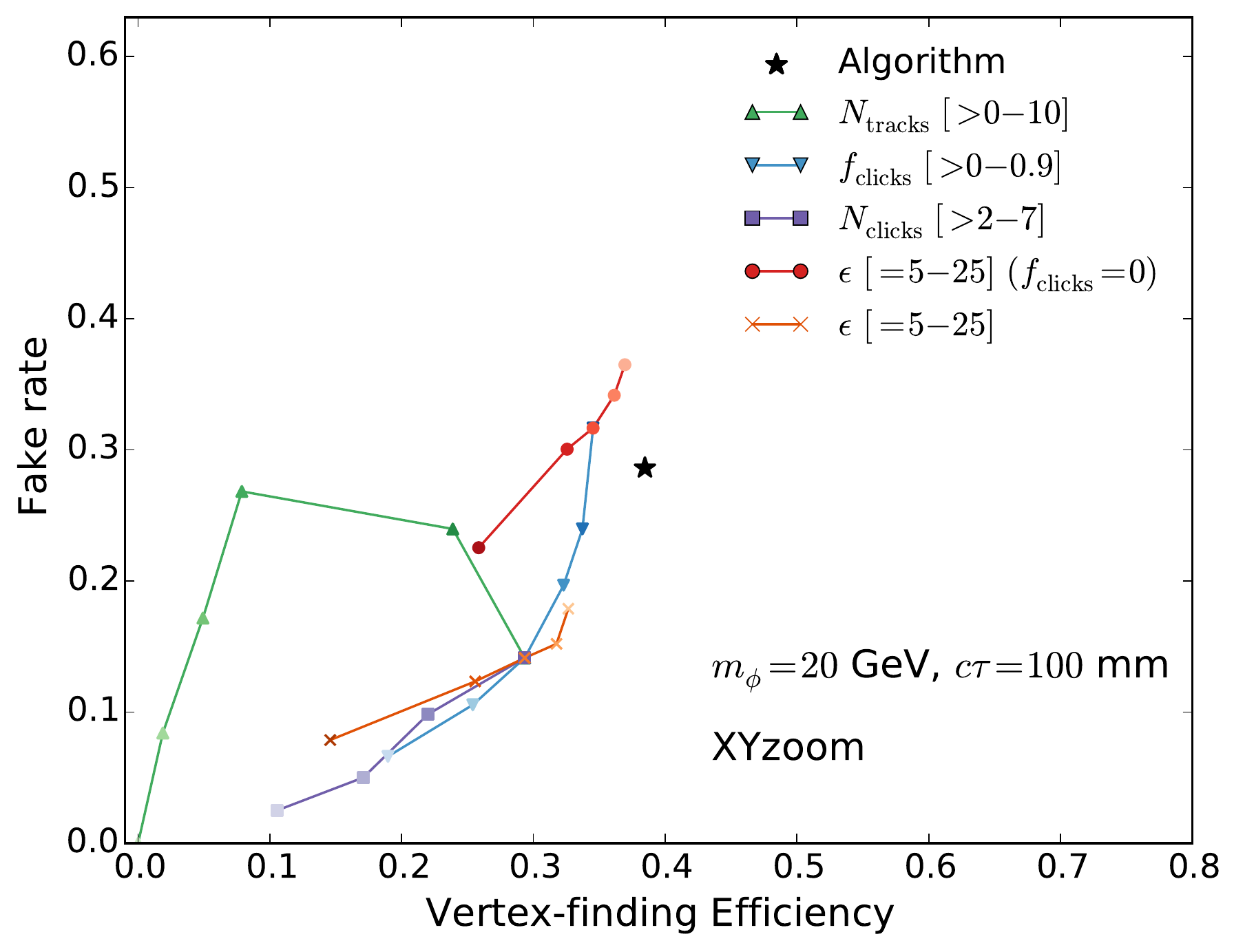}
        \caption{$m_\phi$ = 20 GeV, $c\tauphi$ = 100 mm} 
\end{subfigure}
\caption{Example parameter choices for a selection of simulation types, both in the \texttt{XY} view.}
\label{fig:exampleoptimisation}
\end{figure}

\Cref{fig:exampleoptimisation} shows some examples of how these parameter choices can affect the resulting efficiency and fake rate. 
A baseline choice is taken as $\epsilon = 15$, $N_\textrm{tracks} \geq 0$, $N_\textrm{clicks} \geq 3$, $f_\textrm{clicks} \geq 0.7$, 
striking a balance between high efficiency and low fake rate. 
As a measure comparing the vertices from the reconstruction algorithm and citizen scientists' click clusters, 
the efficiencies for each are shown in \cref{tab:effs}, with clustering parameters chosen so as to give 
the same fake rate for both. 

It can be seen that some $\phi$ parameter sets fare worse, some better and some similarly (e.g. both a lower fake rate and efficiency), 
indicating that the citizen scientists are in many cases outperforming the reconstruction. 


\newcommand\eyewins{\includegraphics[width=7mm,trim=0 400 0 0,clip]{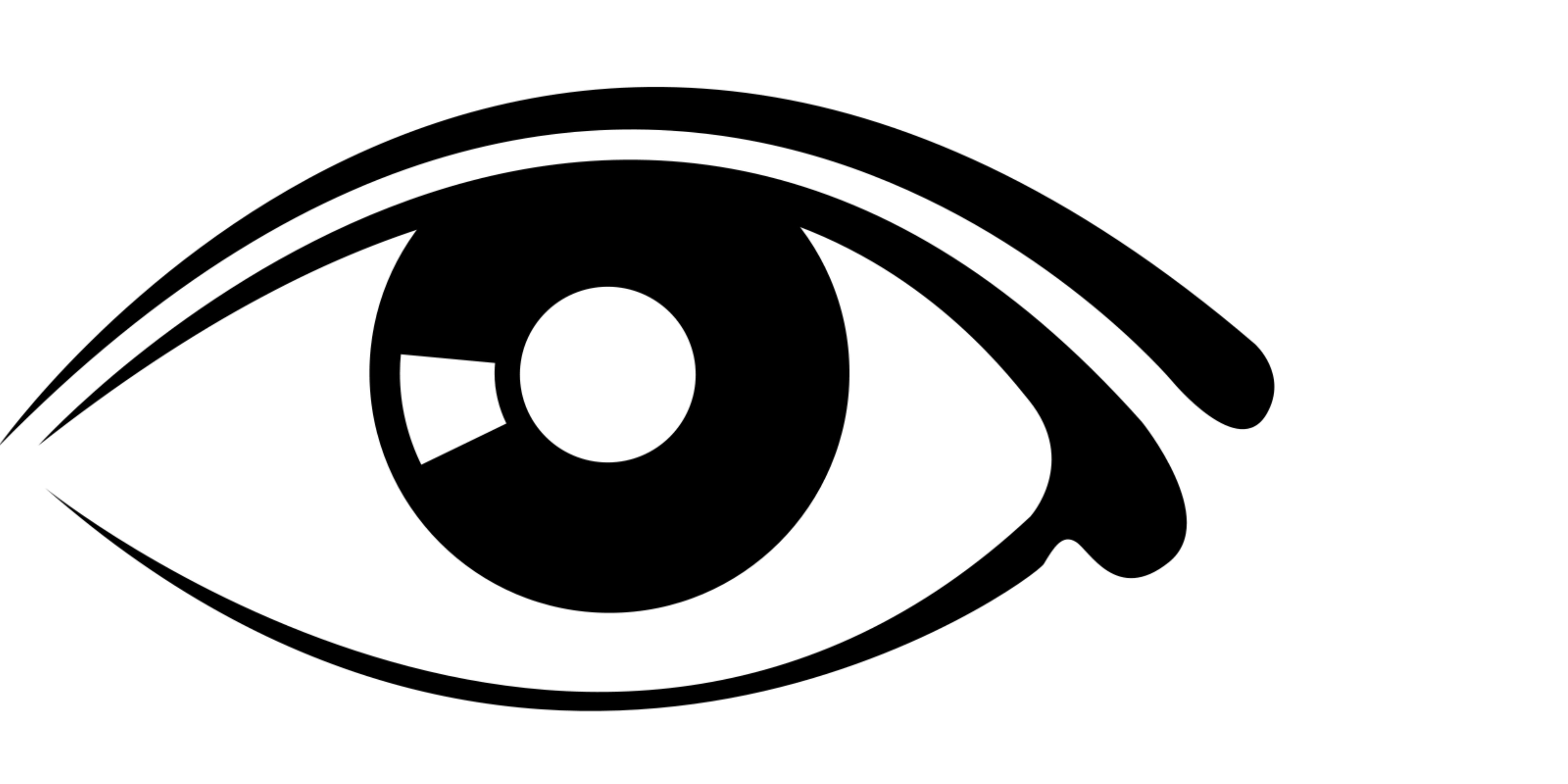}}
\newcommand\computerwins{\includegraphics[width=7mm]{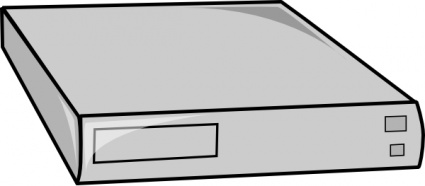}}
\newcommand\closerunthing{
  \includegraphics[width=7mm]{computer}~/~\includegraphics[width=7mm,trim=0 400 0 0,clip]{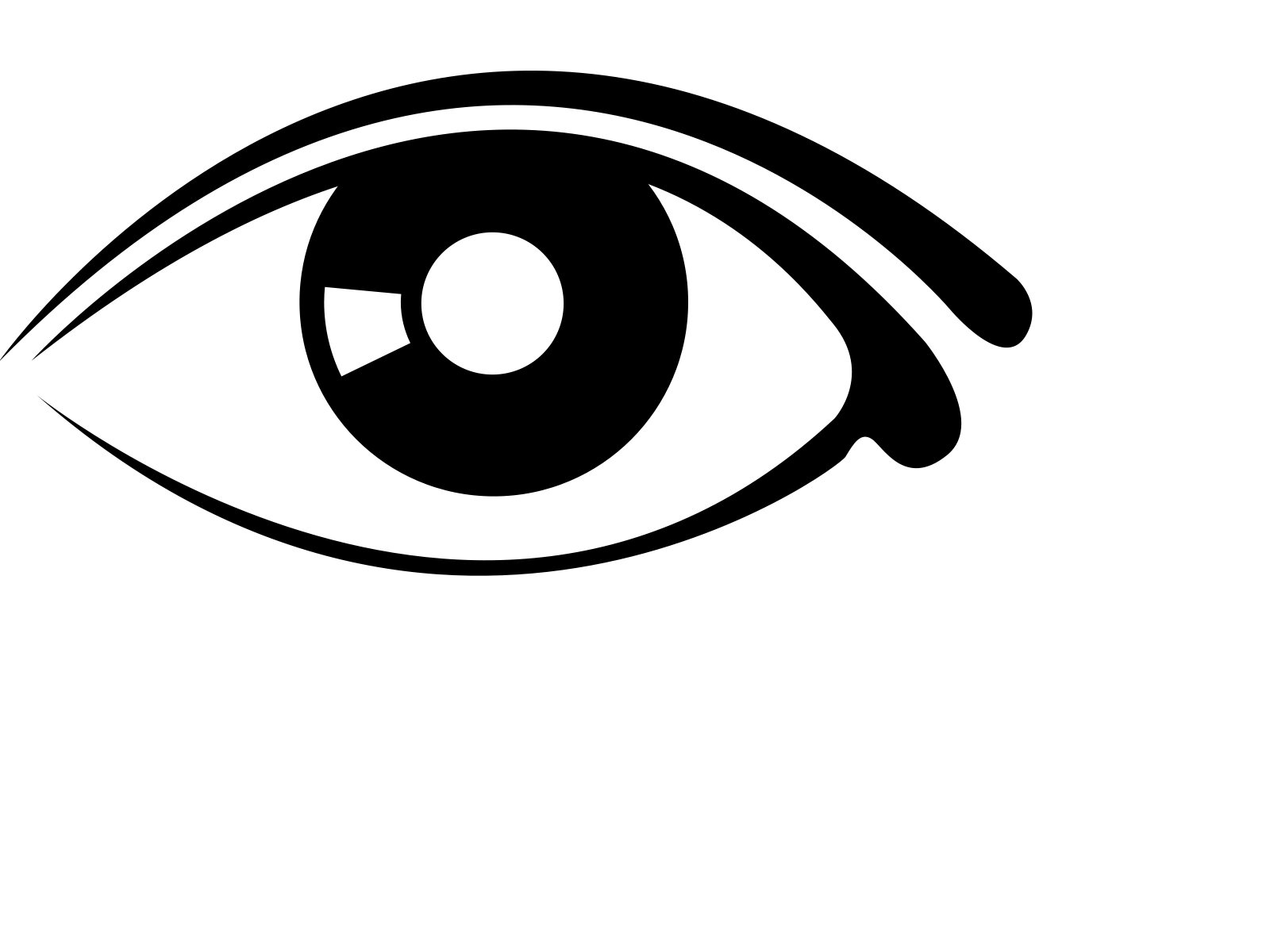}
}
\newcommand\eyejustwins{
\includegraphics[width=7mm,trim=0 400 0 0,clip]{eye-clip-art-eye-biggest}~/~\includegraphics[width=7mm]{computer}
}

\newcommand\eyecolour{} 
\newcommand\compcolour{} 
\newcommand\nearcolour{} 

\begin{table*}
\centering
\renewcommand\arraystretch{1}
\setlength{\tabcolsep}{1.3mm}
\begin{tabular}{| c | c | c | c | c | c | c |}
\hline
\multicolumn{2}{|c|}{$\phi$ properties} & \multirow{2}{*}{View} & 
\multirow{2}{*}{Fake rate [\%]} & \multicolumn{2}{c|}{Efficiency [\%]} & \multirow{2}{*}{Winner} \\
\multicolumn{1}{|c}{$m$ [\GeV]} & $c\tauphi$ [mm]     &      &  & \multicolumn{1}{c}{\computerwins} & \eyewins & \\
\hline
 \eyecolour   8 & 100 & \XY     & 14 &  8  & 14 & \eyewins \\
 \eyecolour   8 & 100 & \XYzoom & 13 &  7  & 12 & \eyewins \\
 \compcolour  8 & 100 & \RZzoom & 12 &  6  & 4 & \computerwins \\
 \eyecolour   8 & 10  & \XY     & 15 & 15  & 27 & \eyewins \\
 \eyecolour   8 & 10  & \XYzoom & 15 & 14  & 29 & \eyewins \\
 \compcolour  8 & 10  & \RZzoom & 12 & 13  & 9 & \computerwins \\
 \eyecolour   8 & 1   & \XY     & 22 & 8   & 9 & \eyejustwins \\
 \eyecolour   8 & 1   & \XYzoom & 21 & 7   & 11 & \eyewins \\
 \eyecolour   8 & 1   & \RZzoom & 16 & 5   & 5 & \eyejustwins \\
 \nearcolour 20 & 100 & \XY     & 27 & 39  & 37 & \closerunthing \\
 \nearcolour 20 & 100 & \XYzoom & 29 & 38  & 34 & \closerunthing \\
 \compcolour 20 & 100 & \RZzoom & 27 & 33  & 21 & \computerwins \\
 \nearcolour 20 & 10  & \XY     & 40 & 59  & $\geq$47 & \closerunthing \\
 \nearcolour 20 & 10  & \XYzoom & 44 & 57  & $\geq$52 & \closerunthing \\
 \compcolour 20 & 10  & \RZzoom & 40 & 56  & $\geq$34 & \computerwins \\
 \compcolour 20 & 1   & \XY     & 26 & 36  & 11 & \computerwins \\
 \compcolour 20 & 1   & \XYzoom & 31 & 34  & 17 & \computerwins \\
 \compcolour 20 & 1   & \RZzoom & 27 & 34  & 12 & \computerwins \\
 \nearcolour 50 & 100 & \XY     & 41 & 59  & $\geq$46 & \closerunthing \\
 \nearcolour 50 & 100 & \XYzoom & 43 & 59  & $\geq$48 & \closerunthing \\
 \compcolour 50 & 100 & \RZzoom & 39 & 57  & $\geq$32 & \computerwins \\
 \compcolour 50 & 10  & \XY     & 51 & 72  & $\geq$35 & \computerwins \\
 \compcolour 50 & 10  & \XYzoom & 53 & 70  & $\geq$41 & \computerwins \\
 \compcolour 50 & 10  & \RZzoom & 50 & 69  & $\geq$28 & \computerwins \\
 \compcolour 50 & 1   & \XY     & 27 & 41  & 9 & \computerwins \\
 \compcolour 50 & 1   & \XYzoom & 31 & 40  & 6 & \computerwins \\
 \compcolour 50 & 1   & \RZzoom & 28 & 40  & 6 & \computerwins \\
\hline
\end{tabular}
\caption{
Vertex-finding efficiencies for citizen scientists (marked with an eye) and the computer algorithm (marked with computer).
Efficiencies are shown for each of the mass and lifetime of the \target{} boson $\phi$ and for the three views considered.
The symbols in the final column are an eye where citizen scientists outperform the algorithm, 
and a computer where the algorithm outperforms.
Where the losing strategy achieves an efficiency of at least 75\% of the winning one (i.e the
result is marginal) its symbol follows that of the winner.
When the fake rate of the algorithm is always higher than that of the citizen scientists
for any set of clustering parameters
then only a lower bound can be placed on the eye efficiency value 
that would be found for equal fake rates.
}
\label{tab:effs}
\end{table*}

Table~\ref{tab:effs} shows a comparison between the algorithm's efficiency and that of 
the citizen scientists for the different masses and lifetimes of the \target{} boson $\phi$.

Since the parameters of the clustering algorithm can be varied to increase vertex-finding 
efficiency as the cost of increasing the number of fakes, 
the efficiency for the citizen scientist clustering is taken as the maximal efficiency 
which gives the same fake rate as obtained by the algorithm. 
This is the rightmost intercept of the coloured lines and a horizontal one through the star in \cref{fig:exampleoptimisation}. 
Where no such intercept exists because the algorithm always has a higher fake rate than the citizen 
scientists for all clustering parameters, the efficiency for the citizen scientists is taken to be that
with the highest fake rate found, and shown as a lower bound.
This is a conservative estimate of their efficiency since it is known that
the citizen scientists can achieve at least this efficiency even at a lower fake rate.

The efficiencies and fake rates differ between views since some vertices fall 
outside the image boundaries and are not counted towards the calculations, 
and not all events have been examined by at least three people in all three views 
(so the set of events considered differs slightly between views). 
In general the citizen scientists had more difficulty 
locating the vertices in the \RZzoom{} view.
This was expected by the experimenters, 
nevertheless the view was included since it offers the possibility of 
three-dimensional vertex reconstruction.

Overall it can be seen that the performance of the citizen scientists competes
very well with that of the computer algorithm.
The collective ability of the citizen scientists 
tends to beat the computer algorithm for simulations with low mass (8\,GeV) \target{} bosons.
This is true for almost all views and all tested values of the lifetimes.
As the mass of the boson $\phi$ increases 
the citizen scientists generally retain a very respectable efficiency, 
but the balance shifts in favour of the algorithm. 

It's interesting to speculate on why the relative performance of 
the algorithm versus the eye depends on the mass of the \target{} boson.
One reason may be that heavier $\phi$ particles have lower velocities $\beta$,
and hence lower Lorentz factors $\gamma$ in the Higgs boson rest frame.
This decreases their travel distance $\beta\gamma c \tauphi$ for any particular $\tauphi$. 
Another factor might be that the heavier $\phi$ particles decay to $b+\bar{b}$ quarks
rather than to $\tau^+ + \tau^-$ leptons as is the case for the lighter $\phi$ bosons.

\section{The `Muon-Jet' Event}
\label{sec:analysis:weird}

In addition to being able to mark off-centre vertices, the citizen scientists are also encouraged to select anything `weird' in the images, 
and to follow up these on the `talk' forum~\cite{HiggsHuntersTalk} 
where the wider community discusses them. 
This raised several instances of known phenomena, such as cosmic ray showers passing through ATLAS, but also some that were unexpected. 

\begin{figure*}
\centering
\includegraphics[width=0.6\textwidth]{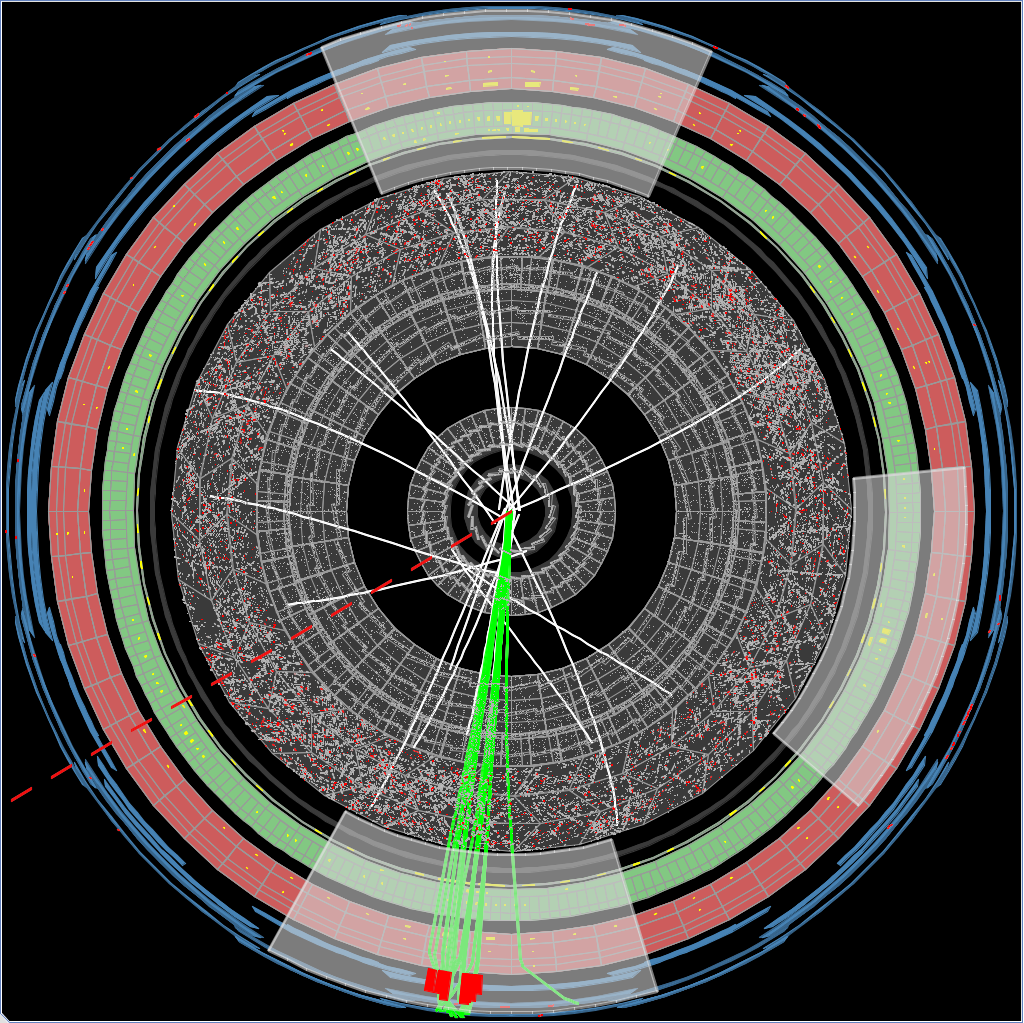}
\caption{The original `muon jet' image, identified by a user on 27th November 2014. 
There appear to be a large number of muons (green lines passing all the way through the detector) all very close together.}
\label{fig:muonjet}
\end{figure*}

Some such oddities were particularly surprising. Soon after the project's launch, the image in \cref{fig:muonjet} 
was flagged as weird and posted in the “talk” forum by several citizen scientists. 
The image shows a collision apparently containing a jet of multiple collimated muons. 
Such a feature is not expected in the Standard Model of particle physics, 
where jets are always of hadrons, not of muons. 
The observations caused a flurry of activity amongst both the citizen scientists and the science team. 

After further investigation by the science team, it was revealed that this event was an example of `punch-through’. 
This is an unusual interaction of a known particle with the detector, rather than an unusual or new particle. 
So while it does not represent a discovery of new physics, 
this example clearly shows the potential for untrained citizen scientists to isolate interesting features in real LHC collision data.

\section{Conclusion}
\label{sec:conclusion}

The first mass participation citizen science project for the Large Hadron Collider has been extremely successful. 
More than \nvolunteers{} citizen scientists participated, and 
more than 1.2 million features of interest were
identified in images from the ATLAS detector.

The collective ability of the citizen scientists was found to be very high. 
They were effective at locating secondary vertices of long-lived particles, 
with efficiency and false-identification rates competitive with (and in some cases better than) 
the ATLAS reconstruction algorithms. 

Amongst the unusual features volunteers spotted were what appeared to be jets of muons, features
unexpected in the Standard Model of particle physics, and which later analysis showed 
to be a feature known as calorimeter punch-through.



\section*{Acknowledgements}

The \url{HiggsHunters.org} project is a collaboration between the University of Oxford and the University of Birmingham in the United Kingdom, 
and NYU in the United States.  It makes use of the \href{www.zooniverse.org}{Zooniverse} citizen science platform, which hosts over 40 projects from searches for new 
astrophysical objects in telescope surveys to following the habits of wildlife in the Serengeti. The HiggsHunters project shows collisions 
recorded by the ATLAS experiment and uses software and display tools developed by the ATLAS collaboration. 
The authors gratefully acknowledge the generous financial support of the UK Science and Technology Facilities Council, the University of Oxford, and Merton College, Oxford.

The data used in this work is available upon request from the Institute for Research in Schools\cite{iris}.

\clearpage
\appendix
\part*{Appendix}
\addcontentsline{toc}{part}{Appendix}

\section{Simulation and image reconstructon details}
\label{sec:simul}

Calibration $Z (\to \mu^+ \mu^-) + H (\to \phi \phi)$ events were generated using \texttt{MadGraph}-5.1.5.2~\cite{MadGraph5} 
interfaced to \texttt{Pythia}-8.175 \cite{pythia6,pythia8} 
using the \texttt{AU2} tune~\cite{Pythia8tunes} of \texttt{Pythia} parameters with the \texttt{CTEQ6L1}~\cite{CTEQ6} 
PDF set. The $\phi$ is a pseudoscalar boson, i.e. a spinless particle with negative parity.



They were passed through ATLAS simulation infrastructure~\cite{SOFT-2010-01}, 
and simulated pileup events from the same and surrounding bunch crossings were added, 
along with modelled detector noise, corresponding to the same luminosity profile as the 2012 data sample.

Tracks and vertices were reconstructed as described in Ref.~\cite{Aad:2015rba}.
In particular, the tracking was extended from 
the ATLAS default impact parameter of 1\,cm, to\,10 cm. 
The total simulation and reconstruction time was about 10\,minutes per event, 
or $\sim$1 CPU-year. (This was spread out to many computers on the ATLAS grid
to process $\sim$50k events in $\sim$2 days.) 

The total data size of $\sim$250\,GB was then reduced to $\sim$100 GB of images. 
These images were made using the ATLAS Atlantis event display~\cite{atlantis,Taylor:2005wi}.
Selections were applied to reduce the amount of visual information (clutter) 
shown in each image, while still allowing vertices to be seen. 
Only tracks with $\pt>2$~\GeV and starting $> 0.5$~mm in the transverse plane from
the beamline were shown (since there are $\sim$1000's of low \pt, tracks originating from the interaction point.) 
Tracks that start more than 20\,cm from the center 
of the detector or the selected collision point in the direction along 
the beamline are not drawn. 
To reduce the number of fake tracks, they must have at least seven 
hits in the silicon strip tracker. Vertices shown must have at least three 
tracks. Muon tracks must have $\pt > 10$~\GeV, and 
jets of hadrons must have $\pt > 40$~\GeV. 
Other objects (photons, electrons, bottom quark jets, etc.) are drawn as
long as they have $\pt > 5$~\GeV.

The images were displayed using a non-linear 
radially dependent fish-eye transform of the form
\[
r^\prime = \frac{a r}{m(1 + c r)}
\]
where $a$, $c$ and $m$ are real positive constants,
and $r$ is the radius in pixels in the $x-y$ view.
This transform serves to increase viewers' attention to and precision within
the central tracking layers of the ATLAS detector.

\bibliography{references.bib,ATLAS.bib,CMS.bib}{}
\bibliographystyle{vancouver}

\end{document}